\documentclass[aps,prb,reprint,amsmath,amsfonts,amssymb,superscriptaddress]{revtex4-2}
\usepackage{graphicx,float,calc}
\usepackage{color,bm}
\usepackage{ulem}
\usepackage{braket}

\usepackage[colorlinks,urlcolor=blue,citecolor=blue,linkcolor=blue]{hyperref}
\begin{document}
\title{Discrete time crystals enabled by Floquet strong Hilbert space fragmentation}

\author{Ling-Zhi Tang}\email{tanglingzhi@quantumsc.cn}
\affiliation{Quantum Science Center of Guangdong-Hong Kong-Macao Greater Bay Area (Guangdong), Shenzhen 518045, China}

\author{Xiao Li}
\affiliation{Department of Physics, City University of Hong Kong, Kowloon, Hong Kong SAR, China}

\author{Z. D. Wang}
\affiliation{Key Laboratory of Atomic and Subatomic Structure and Quantum Control (Ministry of Education), Guangdong Basic Research Center of Excellence for Structure and Fundamental Interactions of Matter, South China Normal University, Guangzhou 510006, China}
\affiliation{Guangdong-Hong Kong Joint Laboratory of Quantum Matter, Frontier Research Institute for Physics, School of Physics, South China Normal University, Guangzhou 510006, China}
\affiliation{Quantum Science Center of Guangdong-Hong Kong-Macao Greater Bay Area (Guangdong), Shenzhen 518045, China}
\affiliation{HK Institute of Quantum Science $\&$ Technology, The University of Hong Kong, Pokfulam Road, Hong Kong, China}

\author{Dan-Wei Zhang}\email{danweizhang@m.scnu.edu.cn}
\affiliation{Key Laboratory of Atomic and Subatomic Structure and Quantum Control (Ministry of Education), Guangdong Basic Research Center of Excellence for Structure and Fundamental Interactions of Matter, South China Normal University, Guangzhou 510006, China}
\affiliation{Guangdong-Hong Kong Joint Laboratory of Quantum Matter, Frontier Research Institute for Physics, School of Physics, South China Normal University, Guangzhou 510006, China}
\affiliation{Quantum Science Center of Guangdong-Hong Kong-Macao Greater Bay Area (Guangdong), Shenzhen 518045, China}

\begin{abstract}
Discrete time crystals (DTCs) are non-equilibrium phases of matter that break the discrete time-translation symmetry and are characterized by a robust subharmonic response in periodically driven quantum systems. Here, we explore the DTC, stabilized by the Floquet strong Hilbert space fragmentation, in a disorder-free periodically kicked XXZ spin chain. We numerically show the period-doubling response of the conventional DTC order, and uncover a multiple-period response with beating dynamics due to the coherent interplay of multiple $\pi$-pairs in the Floquet spectrum of small-size systems. The lifetime of the DTC order exhibits independence of the driving frequency and a power-law dependence on the ZZ interaction strength. It also grows exponentially with the system size {up to a critical imperfection}, as a hallmark of the strong fragmentation inherent to the Floquet model. We analytically reveal the approximate conservation of the magnetization and domain-wall number in the Floquet operator for the emergent strong fragmentation, which is consistent with numerical results of the dimensionality ratio of symmetry subspaces. The rigidity and phase regime of the DTC order are identified through finite-size scaling of the Floquet-spectrum-averaged mutual information, as well as via dynamical probes. Our work establishes the Floquet Hilbert space fragmentation as a disorder-free mechanism for sustaining nontrivial temporal orders in out-of-equilibrium quantum many-body systems.
\end{abstract}

\date{\today}

\maketitle

\section{Introduction}

The notion of time crystals, which exhibit spontaneous breaking of continuous time-translation symmetry, was initially introduced by Wilczek \cite{Wilczek2012}. 
Subsequent studies, however, have demonstrated that such symmetry breaking cannot occur in equilibrium or in the ground states of quantum systems \cite{Bruno2013,Watanabe2015}. 
This limitation was overcome with the introduction of discrete time crystals (DTCs) in periodically driven systems, where the discrete time-translation symmetry is broken in a non-equilibrium manner \cite{Sacha2015,Khemani2016,Else2016,Yao2017,Zaletel2023Review,Khemani2019}.
A key challenge in realizing stable DTCs lies in the generic tendency of Floquet systems to heat up indefinitely, eventually reaching a featureless thermal state \cite{Pedro2015}.
To circumvent this unbounded heating and sustain the DTC order, various mechanisms that inhibit ergodicity have been explored in recent years \cite{Yao2017,Kshetrimayum2020,Kshetrimayum2021,Liu2023,Tang2021,Tang2025,Li2024,Zhang2020,Liang2023,Ni2021,JJWang2025,Yousefjani2025}.
A prominent route exploits many-body localization induced by strong disorders \cite{Abanin2019,Sierant2025}. 
Alternative avenues involve prethermalization \cite{Else2017,Pizzi2021,kyprianidis2021,Luitz2020,Stasiuk2023,Beatrez2023,Hu2023,CYing2022,ZHBao2025} or the phenomenon of quantum many-body scars, which restrict thermalization for specific initial states \cite{Maskara2021,Huang2022,Huang2024,Kumar2025}. 
To date, experimental signatures of DTCs have been reported in various engineered quantum platforms \cite{Choi2017,Zhang2017,Rovny2018PRL,Rovny2018PRB,Autti2018,Pal2018,Smits2018,Giergiel2020,Smits2021,Mi2022,Frey2022,Stasiuk2023,Beatrez2023,CYing2022,Liu2024,Wang2021,DeRoeck2017,Ho2017,DYZhu2025,GHHe2025}, such as cold atoms \cite{DYZhu2025}, trapped ions \cite{Zhang2017}, superconducting qubits \cite{Mi2022,Frey2022,ZHBao2025}, dipolar spin ensembles \cite{Choi2017}, nitrogen-vacancy centers \cite{GHHe2025}, and nuclear magnetic resonance setups \cite{Rovny2018PRL}.

While many-body localization provides a robust and initial-condition-independent mechanism for protecting DTCs, its characterization often necessitates averaging over numerous disorder realizations, which poses a severe computational burden in simulations and consumes considerable resources in experimental settings.
This has motivated the exploration of alternative nonergodic mechanisms for realizing DTCs in disorder-free settings \cite{Russomanno2017, Gargiulo2024, Else2017,Pizzi2021,kyprianidis2021,Luitz2020,Stasiuk2023,Beatrez2023,CYing2022,ZHBao2025,JJWang2025,Yousefjani2025}. 
A particularly promising direction is Hilbert space fragmentation (HSF), a phenomenon where a system's Hilbert space fragments into numerous dynamically disconnected subspaces \cite{Sala2020a,Khemani2020a}. 
This fragmentation can stem from kinetic constraints \cite{Khemani2020a,Sala2020a,Pai2019a,LZhang2024} or restricted local tunneling \cite{DeTomasi2019a,ZCYang2020,PFrey2022,SGhosh2023}. 
Experimentally, HSF has now been observed in systems ranging from superconducting processors \cite{YYWang2025} to cold atoms \cite{Scherg2021,Kohlert2023,Adler2024,Will2024,Honda2025,LHZhao2025}, with recent extensions to two-dimensional geometries \cite{Adler2024,Will2024}. 
Most of these studies focus on static systems, while the HSF in out-of-equilibrium quantum systems remains largely unexplored \cite{Kohlert2023}. Moreover, its potential for stabilizing DTCs in Floquet systems is yet to be studied.

In this work, we reveal a disorder-free DTC order enabled by Floquet strong HSF in a periodically kicked XXZ spin chain. Apart from the period-doubling response as the hallmark of the DTC, we find that the system can exhibit multiple-period response with long-period beating dynamics for domain-wall initial states in the small-size case, which originates from the coherent interplay of multiple $\pi$-pairs in the quasi-energy spectrum. We numerically characterize the lifetime of the DTC order, which is independent of the driving frequency, follows a power-law increasing with the ZZ interaction strength, and grows exponentially with the system size as a hallmark of the underlying strong fragmentation. We analytically show the approximate conservation of the magnetization and domain-wall number in the Floquet operator for the emergence of strong HSF under strong ZZ interactions, which is consistent with numerical results of the structure and dimension of symmetry subspaces. Furthermore, we study the rigidity of the DTC against driving imperfections, and obtain its phase regime using both static and dynamical probes. Given its disorder-free nature, the DTC enabled by the Floquet HSF is suitable for realization on near-term quantum simulators. Our work also highlights the potential of the HSF for stabilizing and exploring non-equilibrium quantum phases in disorder-free many-body systems.

The rest of the paper is organized as follows. Sec.~\ref{sec:model} introduces the Floquet spin model to realize disorder-free DTCs. In Sec.~\ref{sec:results and dicussion}, we numerically show the dynamical response and lifetime of the DTC order and analyze its protection by the Floquet HSF mechanism. The rigidity and phase regime of the DTC 
order are also studied. Finally, a short conclusion is presented in Sec.~\ref{sec:conclusion and outlook}.

\section{Floquet model} \label{sec:model}

We consider a spin-1/2 chain of length $L$ under the Floquet drive that alternates between two types of Hamiltonian dynamics. The first evolution is a global $\pi$-pulse around the $\hat{x}$ axis with an imperfection $\epsilon$. The second one is under a disorder-free XXZ model for time $T$. The driven dynamics is described by the Floquet operator $U_F=U_2 U_1$ with two evolution operators
\begin{equation}
    U_1= e^{-i(\frac{\pi}{2}-\epsilon)\sum_{j}^L  \sigma^x_j},~~U_2 = e^{-iTH}
\end{equation}
with

\begin{equation}\label{XXZmodel}
    H = \sum_{j}^{L-1} 2 J\left( \sigma _{j}^{x}\sigma _{j+1}^{x}+\sigma _{j}^{y}\sigma _{j+1}^{y} \right) + V\sigma _{j}^{z}\sigma _{j+1}^{z}.
\end{equation}
Here $\sigma _{j}^{u}$ is the $u$-th ($u=x,y,z$) component of the Pauli operator for the $j$-th spin, and we adopt the convention $\hbar=1$. The parameters $J$ and $V$ denote the XY and ZZ interaction strengths, respectively. 
Hereafter, we set $J=1$ as the energy unit. The Floquet operator
\begin{equation}
U_F=e^{-iH_FT} 
\end{equation} 
implements the dynamics over a period of $T$ and has the frequency $\omega=2\pi/T$, where $H_F=\frac{i}{T}\ln U_F$ is the corresponding effective Floquet Hamiltonian. From the eigen-equation 
\begin{equation}
U_F \ket{\psi^{F}_\alpha}= e^{-i\varepsilon_{\alpha}T} \ket{\psi^{F}_\alpha}, 
\end{equation}
one can obtain Floquet eigenstates $\ket{\psi^F_{\alpha}}$ and the folded quasi-energies $\varepsilon_{\alpha}\in[-\pi/T,\pi/T]$ with $\alpha=\{1,2,..,D\}$, where $D=2^L$ is the Hilbert-space dimension of the system. Note that this Floquet model could be realized in several quantum simulators based on superconducting qubits \cite{Mi2022,Frey2022,XZhang2022,CYing2022,ZHLiu2025,Will2025}, Rydberg atom arrays \cite{Bluvstein2021,Kim2024,Koyluoglu2025,Evered2025} and trapped-ion qubits \cite{Zhang2017,kyprianidis2021,Dumitrescu2022}

In the absence of the XY interaction term ($J=0$), the model reduces to the binary Floquet Ising chain studied in Refs. \cite{Yao2017,Else2016,Keyserlingk2016,ZHBao2025}. 
In this case, we can simply analyze the subharmonic response as the Floquet time crystalline order under an ideal $\pi$-pulse with $\epsilon=0$. 
To see this, we consider an arbitrary initial product state aligned along the $\hat{z}$ axis undergoing the spin-echo evolution \cite{Yao2017}. The first unitary $U_1=-i\prod_j\sigma _{j}^{x}$ flips each spin about the $\hat{x}$ axis and results in the oppositely polarized state. Since each spin is already along the $\hat{z}$ axis, the second unitary $U_2=e^{-iTV\sum_{j}\sigma _{j}^{z}\sigma _{j+1}^{z}}$ only gives rise to a global phase $\phi=TV\sum_{j}s_js_{j+1}$, where $s_j=\pm1$ are eigenstates of $\sigma _{j}^{z}$. 
As each spin is flipped once per Floquet period, the magnetization response [such as the fidelity in Eq. (\ref{F}) and the autocorrelation in Eq. (\ref{C})] at stroboscopic times $t=nT$ ($n=1,2,...$) yields a perfect $2T$-periodic oscillation. This corresponds to the subharmonic Fourier response at half of the drive frequency and characterizes the time crystalline order. 
However, imperfections ($\epsilon>0$) in the drive will destroy the subharmonic response and melt the DTC to a thermal phase after long-time evolution. 
Thus, localization induced by strong disorders in an additional longitudinal field (and the Ising interaction) is crucial to stabilize the DTC phase, as shown in Ref. \cite{Yao2017}. 
For our model without disorders, the XY interaction ($J=1$) is also expected to destroy the time crystals response as it results in the transport of magnetization excitations. 
In the following, we show {that} the time {crystalline} order can persist for long times in this disorder-free system, which is enabled by the Floquet HSF under strong ZZ interaction.

\section{Results} \label{sec:results and dicussion}

In this section, we investigate the DTC dynamics in our Floquet model by numerically simulating the time evolution from {the} initial product state $|\psi(0)\rangle$. The wavefunction at stroboscopic times $t=nT$ is given by $|\psi(t)\rangle=U_{F}^{n}|\psi(0)\rangle$. The DTC response in our model falls into two distinct categories based on their frequency signatures. To illustrate each type, we examine two canonical initial states: the Néel state and the domain-wall state, whose dynamics capture the essential features of the two types of DTC. As the model is generally non-integrable, we employ exact diagonalization \cite{Weinberg2019} for systems with $L\leq 14$ and the time-evolving block decimation \cite{Fishman2022} for larger systems with $24\leq L\leq60$ under open boundary conditions. The static properties related to the effective Floquet Hamiltonian $H_{F}$ are also analyzed using exact diagonalization.

\begin{figure}[tb]
	\centering
	\includegraphics[width=0.48\textwidth]{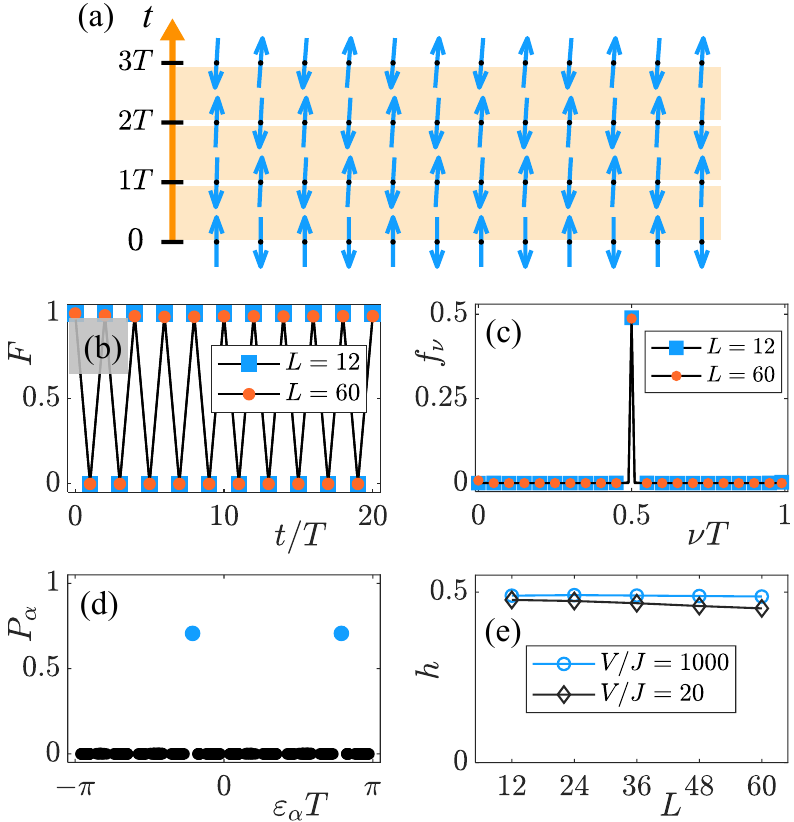}
	\caption{(Color online) Period-doubling DTC order for a N\'eel initial state. 
	(a) Spatiotemporal spin configuration. 
	(b) Evolution of fidelity $F(t)$ for $L = 12,60$. 
	(c) Fourier spectrum $f_\nu$, showing a prominent peak at the subharmonic frequency $\nu = 1/2T$. 
	(d) Overlap between the initial state and Floquet eigenstates $P_\alpha$. Here, the dominant $\pi$-pair is marked by blue squares. 
	(e) System-size dependence of the subharmonic amplitude $h$ for $V/J = 20,1000$. 
	Other parameters: $L=12$ in (a,d), $V/J = 1000$ in (a-d), $\epsilon = 0.01$ and $T=1$ throughout.}
	\label{fig1}
\end{figure}

\begin{figure}[tb]
	\centering
	\includegraphics[width=0.48\textwidth]{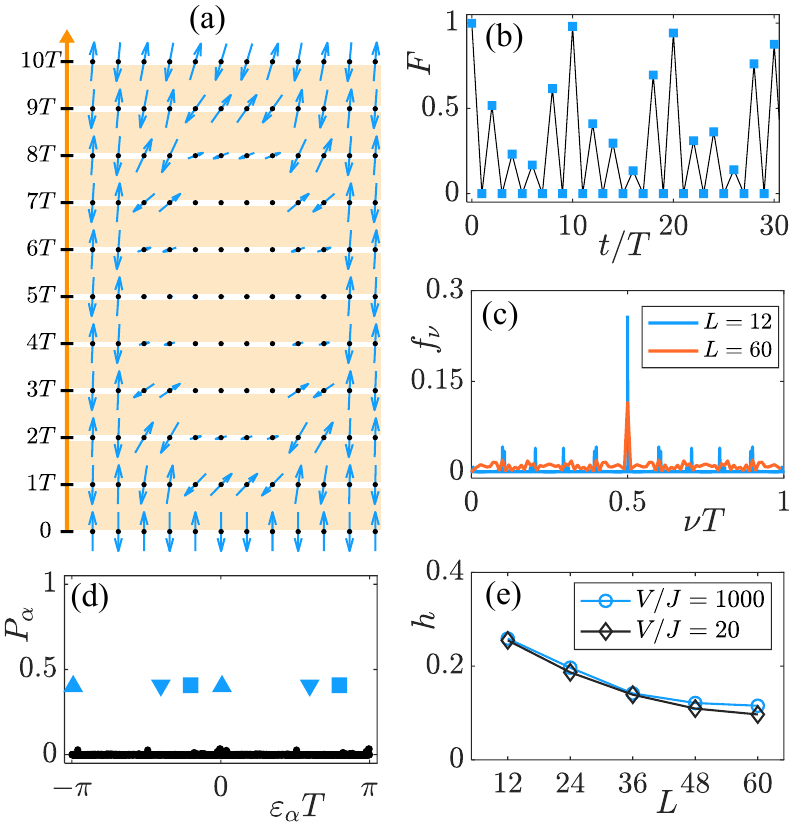}
	\caption{(Color online) \label{fig2}
	Multi-period DTC order for a domain-wall initial state. 
	(a) Spatiotemporal spin configuration. 
	(b) Evolution of fidelity for $L = 12$. 
	(c) Fourier spectrum $f_\nu$ for $L = 12$ and $L = 60$. 
	(d) Overlap $P_\alpha$ between the initial state and all Floquet eigenstates for $L=12$. Symbols distinguish three $\pi$-pairs. 
	(e) Subharmonic amplitude $h$ versus $L$ for $V/J = 20, 1000$. 
	Other parameters: $L=12$ in (a,d), $V/J = 1000$ in (a-d), $\epsilon = 0.01$ and $T=1$ throughout.
	}
\end{figure}

{\subsection{DTC responses}}

We first study the Floquet dynamics starting from a N\'{e}el state $\ket{\psi(0)}=\ket{\uparrow\downarrow\uparrow\downarrow\cdots}$ under strong ZZ interactions $V/J=1000,20$ and small imperfection $\epsilon=0.01$. A spatiotemporal configuration of the spin chain of length $L=12$ is numerically obtained in Fig.~\ref{fig1}(a), where a $2T$-period can be seen from the spin orientation. To further diagnose {the} period-doubling response of the DTC order, we compute the time evolution of the fidelity
\begin{equation}\label{F}
	F(t)=|\braket{\psi (0) | \psi(t)}|.
\end{equation}
As shown in Fig.~\ref{fig1}(b), the fidelity for two systems of lengths $L=12,60$ both exhibit the period-doubling dynamics. {To quantify these dynamics}, we calculate the fast Fourier transform of $F(t)$ in the long-time evolution: $f_\nu = \sum_{0}^{N_t} F(nT) e^{- i 2 \pi n \nu}$, where $\nu$ denotes the response frequency and $N_t=500$ is taken in our numerical simulations. 
Figure \ref{fig1}(c) depicts a prominent Fourier peak at the subharmonic frequency $\nu=1/2T$ with the amplitude $h\equiv f_{\nu=1/2T}\approx0.5$. Such a subharmonic response, corresponding to a $2T$-period, is the hallmark of the DTC order. Note that this period-doubling response is independent of the driving period $T$ (frequency $\omega$) and can be exhibited starting from other initial states. 
We also observe similar frequency patterns [See Fig.~\ref{fig1}(c)] for other initial states that are only dynamically connected to their own spin‑reversed states.
Examples include $\ket{\uparrow\uparrow\uparrow\uparrow\downarrow\downarrow\downarrow\downarrow}$, $\ket{\uparrow\uparrow\uparrow\uparrow\uparrow\uparrow\uparrow\downarrow}$ and $\ket{\uparrow\downarrow\downarrow\downarrow\downarrow\downarrow\downarrow\uparrow}$. 
These configurations are termed frozen states, characterized by the absence of movers under the strong kinetic constraints \cite{DeTomasi2019a}.

The subharmonic DTC response is related to the subspace dimension of the initial state under the Floquet basis. To illustrate this point, we calculate the overlap between the initial state and Floquet eigenstates \cite{Liu2023}
\begin{equation}
P_{\alpha}=|\braket{\psi^F_\alpha|\psi (0)}|. 
\end{equation}
As shown in Fig.~\ref{fig1}(d), the spectrum $P_{\alpha}$ with respect to all Floquet eigenstates of quasi-energies $\varepsilon_{\alpha}$ reveals a dominant so-called $\pi$-pair, which is {a} pair of Floquet eigenstates with a quasi-energy splitting of $\pi$. The $\pi$-pair structure confines the system dynamics predominantly to a two-{dimensional} subspace with a period-$2T$ response. 
We numerically extract the amplitude of the subharmonic response $h$ as a function of system size $L$, with the results shown in Fig.~\ref{fig1}(e). 
For strong ZZ interaction of strengths $V/J=1000$ and $20$, the amplitude $h$ remains large (with only a small decay) {for system sizes} up to $L=60$. 
As we show in the next subsection, the subharmonic DTC response exhibits remarkable persistence {as we increase the system size, the ZZ interaction strength and the driving frequency}.

We further study the dynamics beginning from distinct initial states with domain-wall spin configurations, such as$\ket{\psi(0)}=\ket{\uparrow\downarrow\uparrow\downarrow\uparrow\downarrow\downarrow\uparrow\downarrow\uparrow\downarrow\uparrow}$ for $L=12$ in Fig.~\ref{fig2}(a). 
By embedding adjacent spin pairs  (``$\uparrow\uparrow$'' or ``$\downarrow\downarrow$'') as defects into the antiferromagnetic background, we obtain similar domain-wall initial states for $L=24,36,48,60$ with the same domain-wall number. The DTC dynamics of the domain-wall initial states are summarized in Fig.~\ref{fig2}. The spatiotemporal configuration in Fig.~\ref{fig2}(a) shows the complex response dynamics: The edge spins exhibit a period-$2T$ oscillation, while the entire spin chain exhibits a near-perfect recurrence at $t=10T$. This global revival is quantified by a distinct peak in the fidelity dynamics $F(t)$ in Fig.~\ref{fig2}(b), which reveals a multi-period response with a beating period of $T_b=10T$.
The corresponding Fourier spectrum for $L=12$ in Fig.~\ref{fig2}(c) {shows} more clearly the multi-period DTC response. 
Apart from a prominent subharmonic peak at $\nu=1/2T$ with $h\approx0.25$, the spectrum has eight sub-leading peaks at frequencies $\nu=k/10T$ with $k=1,2,\ldots,9$ (excluding $k=5$). 
Thus, a beating period $T_b=10T$ corresponds to the smallest common period encompassing all these frequency components.

The coexistence of these subharmonic frequencies originates from the coherent interplay of multiple $\pi$-pairs, as evidenced by the overlap of the quasi-energy spectrum in Fig.~\ref{fig2}(d), where different symbols mark three $\pi$-pairs for the domain-wall initial state of length $L=12$. The dominated subharmonic frequency $\nu=1/2T$ comes from each $\pi$-pair. The frequencies of other sub-leading peaks $\nu=\nu_F$ are determined by the quasi-energy differences $\delta_F$ among these $\pi$-pairs \cite{Liu2023}:
\begin{equation}
	\label{eq:frequency}
	\nu_F = \frac{\delta_F}{2\pi T}.
\end{equation}
To illustrate this point, one can consider an initial state decomposed into two such $\pi$-pairs with {an} energy difference $\delta_F$ as $\ket{\psi(0)} = \ket{\Psi_1} + \ket{\Psi_2}$, where $|\Psi_{1,2}\rangle$ denotes the superposition of Floquet eigenstates of the two $\pi$-pairs. After an even number of periods $2nT$, the state evolves as $\ket{\psi(2nT)} \approx \ket{\Psi_1} + e^{-i2n \delta_F} \ket{\Psi_2}$. The revival to the initial state occurs when the phase factor $e^{-i2n\delta_{F}}=1$, which yields the frequency relation in Eq. \eqref{eq:frequency}. By numerically extracting all quasi-energy differences $\delta_F$ for the three $\pi$-pairs in Fig.~\ref{fig2}(d), we confirm the eight sub-leading frequencies $\nu_F\approx k/10T$ with $k=1,2,3,4,6,7,8,9$, as {expected} from Eq. \eqref{eq:frequency}.

Notably, we find that this multi-period DTC response is susceptible to finite-size systems. 
By increasing the system size up to $L=60$, the amplitude of the primary subharmonic peak is $h\approx0.11$, while the sub-leading peaks become obscure, as shown in Fig.~\ref{fig2}(c). The subharmonic amplitude $h$ against the system size $L$ for $V/J=1000,20$ is plotted in Fig.~\ref{fig2}(e). In both cases, $h$ exhibits a decrease as $L$ approaches 60. This can be attributed to the growth of the dimension of symmetry subspaces with respect to the domain-wall numbers (see the next subsection). The subharmonic amplitudes of the DTC order in  Figs.~\ref{fig2}(e) and \ref{fig1}(e) seem to {remain} finite when approaching the thermodynamic limit.

\begin{figure}[tb]
	\centering
	\includegraphics[width=0.48\textwidth]{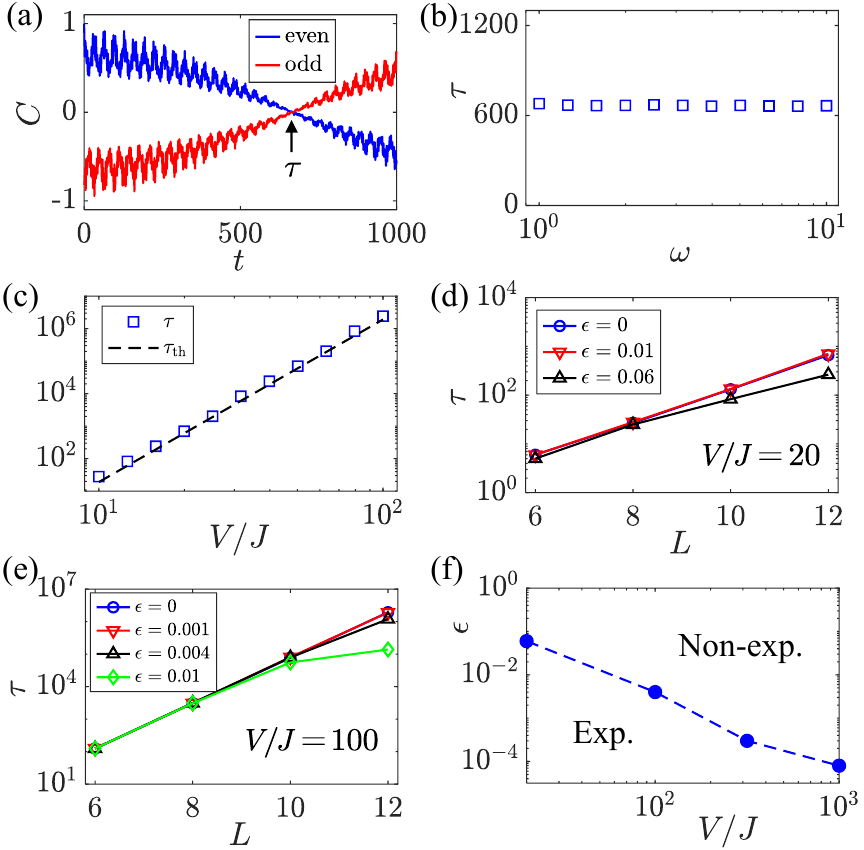}
	\caption{(Color online) \label{fig3}
	Lifetime of the DTC order for domain-wall states. 
	(a) Magnetization autocorrelation $C(t)$, with the lifetime $\tau$ marked for $V/J=20$. 
	(b) $\tau$ versus the driving frequency $\omega$ for $V/J=20$. 
	{(c) $\tau$ and $\tau_{\rm th}$ (given by Eq.~(\ref{tau_th})) versus the ZZ interaction strength $V/J$. 
	(d) System-size scaling of $\tau$ for different drive imperfections $\epsilon$ with $V/J = 20$. (e) Same as (d) but for $V/J = 100$. (f) The extracted critical imperfection $\epsilon_c^\tau$ separates two regimes of exponential (Exp.) and non-exponential (Non-exp.) growths of $\tau$.
	Other parameters: $\epsilon = 0$ in (a-c), $\omega=2\pi$ in (a,c-f).}
	}
\end{figure}

{\subsection{DTC lifetime}}
We proceed to study the lifetime of the DTC order with respect to system parameters.
To quantify the DTC lifetime, we employ the spatially averaged autocorrelation function of the magnetization \cite{Else2016}
\begin{equation}\label{C}
	C(t) = \frac{1}{N_j}\sum_{j} \langle \sigma_j^z(t)\rangle \langle\sigma_j^z(0)\rangle,
\end{equation}
where $N_j=L/2$ as averaging over even and odd sites is respectively computed for odd spin chains. 
As shown in Fig.~\ref{fig3}(a), we plot even- and odd-site $C(t)$ for the domain-wall initial state under strong ZZ interaction strength $V/J=20$. 
One can estimate the lifetime $\tau$ of the period-doubling from the crossover of two lines of $C(t)$, when $t=\tau$ they touch each other for first time. 
Accordingly, we extract the lifetime $\tau$ as a function of the driving frequency $\omega$ for $V/J=20$, as shown in Fig.~\ref{fig3}(b). The lifetime shows negligible variation with frequency, indicating its independence from the driving frequency. This behavior markedly differs from prethermal DTCs, where the lifetime typically exhibits an exponential dependence on the driving frequency \cite{Else2017,Abanin2017}. This conclusion also holds for domain-wall (and other) initial states.
Fig.~\ref{fig3}(c) presents the dependence of the lifetime $\tau$ on the interaction strength $V$, which follows a power-law scaling $\tau/T\propto(V/J)^\mu$ with a fitted exponent $\mu\approx4.96$ (in units of $1/T$).

{To demonstrate the robustness of the DTC order, we introduce a finite drive imperfection $\epsilon$ (as in Fig.~\ref{fig1} and Fig.~\ref{fig2}). Figure \ref{fig3}(d) displays $\tau(L)$ for $V/J = 20$ and different values of $\epsilon$. The exponential growth persists for $\epsilon = 0$ and $\epsilon = 0.01$, but the scaling breaks down once $\epsilon$ reaches $0.06$. From this crossover we extract the critical imperfection $\epsilon_c^\tau \approx 0.06$ below which exponential scaling preserves.
Similarly, critical imperfection $\epsilon_c^\tau \approx 0.004$ can be estimated for $V/J=100$ in Fig.~\ref{fig3}(e). 
Repeating this analysis for additional interaction strengths, we extract $\epsilon_c^\tau(V)$ in Fig.~\ref{fig3}(f), which reveals a rapid decrease of $\epsilon_c^\tau(V)$ with increasing $V/J$. This decrease arises because, as $V/J$ increases, the effective coupling strength between different symmetry subspaces is suppressed more strongly. Consequently, even weaker drive imperfections are sufficient to cause significant subspace leakage and eventual thermalization.
This boundary separates two regimes of exponential and non-exponential growths of $\tau(L)$ with the system size L.

To understand the physical origin of both the exponential scaling of $\tau(L)$ and its sensitivity to drive imperfections, we analyze the high-order virtual processes that couple the initial Néel state $|\!+\rangle$ to its orthogonal anti-Néel partner $|-\rangle$ within the same $\pi$-pair. For the ideal drive ($\epsilon = 0$) and $V \gg J$, these two states can be connected through virtual processes involving exactly $L/2$ applications of the XY terms
\begin{equation}
	H_{XY} = \sum_{j=1}^{L-1} 2J\,(\sigma_j^x\sigma_{j+1}^x + \sigma_j^y\sigma_{j+1}^y).
\end{equation}
Treating $H_{XY}$ as a perturbation on the dominant ZZ part 
\begin{equation}
H_{ZZ} = \sum_{j=1}^{L-1} V\,\sigma_j^z\sigma_{j+1}^z
\end{equation}
of Eq.~(\ref{XXZmodel}), the effective coupling arises at $(L/2)$-th order. The leading contribution is given by the sum over all virtual paths:
\begin{equation}
\begin{split}
	V_{\rm eff} &= \sum_{\rm path} \frac{\langle - |H_{XY}| i_{L/2-1}\rangle \cdots \langle i_2 |H_{XY}| i_1\rangle \langle i_1 |H_{XY}| +\rangle}{\prod_{n=1}^{L/2-1} \bigl(\langle + |H_{ZZ}| +\rangle - \langle i_n |H_{ZZ}| i_n\rangle\bigr)} \\
	&\approx (-1)^{L/2-1} \frac{2^L J^{L/2}}{V^{L/2-1}}.
\end{split}
\end{equation}
The Néel and anti-Néel states thus form an effective two-level system with associated Rabi frequency and period read 
\begin{equation}
	\Omega_{\rm R} = 2|V_{\rm eff}| \ \ \ {\rm and} \ \ \  T_{\rm R} = 2\pi/\Omega_{\rm R},
\end{equation}
respectively. In our work, the lifetime $\tau$ is defined from the crossover of the autocorrelation $C(t)$ (corresponding to a quarter of the Rabi period), hence the theoretical lifetime reads
\begin{equation}\label{tau_th}
	\tau_{\rm th} \approx \frac{T_{\rm R}}{4} = \frac{\pi}{4}\frac{V^{L/2-1}}{2^L J^{L/2}},
\end{equation}
which grows exponentially with $\tau_{\rm th}\propto V^{L/2-1}$ for $V\gg J$. This analytical result is in agreement with the numerically extracted lifetimes shown in Fig.~\ref{fig3}(c). In particular, for the system size $L=12$ used in Fig.~\ref{fig3}(c), the theoretical power-law exponent $L/2-1=5$, which matches the fitted value $\mu\approx 4.96$ to high accuracy.

For small drive imperfection (e.g., $\epsilon = 0.01$ in Fig.~\ref{fig3}(d) and (e)), the imperfect spin-flip operator $U_1$ opens an additional tunneling channel. The minimal process that generates a correction $V_{\rm eff}^\epsilon$ to the effective coupling consists of $(L/2-1)$ applications of the XY terms on the Néel state, followed by two imperfect X operations (each contributing a factor of $\epsilon$). This path therefore appears at least at second order in $\epsilon$. Perturbation theory shows that this correction remains negligible ($V_{\rm eff}^\epsilon \ll |V_{\rm eff}|$) provided $\epsilon$ is sufficiently small ($\epsilon < \epsilon_c^\tau$). Consequently, the original $V_{\rm eff}$ continues to dominate, preserving the exponential growth observed numerically. This high-order perturbative picture provides a direct explanation for the exponential lifetime and sets the stage for the full Floquet HSF analysis in the following subsection.}

\vspace{0.3in}
\subsection{Floquet HSF mechanism}

The mechanism of stabilizing the DTC order in our Floquet model is the emergent symmetry in the Floquet operator $U_F$. In the unperturbed limit ($\epsilon=0$), the driving pulse $U_1$ reduces to a global spin flip $U_1=-i\prod_j \sigma_j^x$, and $U_1$ and $U_2$ conserve the absolute total magnetization $Q = \left| \sum_{j=1}^L \sigma_j^z \right|$. The magnetization $Q$ takes the quantum number $q \in \{0, 2, 4, \ldots, L\}$, which corresponds to the eigenvalues of the symmetry operator.
Here, the absolute value is applied element-wise to the diagonal matrix representation of the total magnetization. This implies that $U_1$, $U_2$, and consequently the full Floquet operator $U_F$ all commute with $Q$: $[U_{1}, Q] = 0$, $[U_{2}, Q] = 0$, $[U_{F}, Q] = [U_{2}U_{1}, Q] = 0$.
The conservation of $q$ fragments the Hilbert space of the Floquet Hamiltonian $H_{F}$ into distinct symmetry sectors $\mathcal{H}_{q}$. The dimension of $\mathcal{H}_{q}$ is given by 
\begin{equation}
	\mathcal{D}_{q} = \frac{a_{q}}{2}\left(\mathcal{C}_{L}^{N_{\uparrow}} + \mathcal{C}_{L}^{N_{\downarrow}}\right), \quad a_{q} = \begin{cases} 1, & q = 0, \\ 2, & q > 0, \end{cases} 
\end{equation}
where $\mathcal{C}$ denotes the binomial coefficient, $N_{\uparrow} = (L + q)/2$ and $N_{\downarrow} = (L - q)/2$ are the numbers of up and down spins, respectively.

A deeper fragmentation arises in the strong ZZ interaction regime $V \gg J$. It is known that such strong constraints lead to HSF in {the} static one-dimensional $t-V$ model \cite{Dias2000}, which can be {mapped} onto the XXZ spin chain under the Jordan-Wigner transformation \cite{Wigner1928a}. In our Floquet spin model with $V \gg J$, the Hamiltonian $H$ in $U_2$ [see Eq. \eqref{XXZmodel}] can be approximated by the projected form~\cite{Dias2000,SGhosh2023}
\begin{equation}
	H \approx H' = \sum_{j=1}^{L-1} \left[ J P_{j} \left( \sigma_{j}^{x} \sigma_{j+1}^{x} + \sigma_{j}^{y} \sigma_{j+1}^{y} \right) P_{j} + V \sigma_{j}^{z} \sigma_{j+1}^{z} \right],
\end{equation}
up to the first order of $J$ (and neglecting corrections of order $J^{2}/V$). Here the projector $P_{j} \equiv 1 - \left( \sigma_{j-1}^{z} - \sigma_{j+2}^{z} \right)^{2}$ enforces a kinetic constraint. The hopping is permitted only if it preserves the total number of domain walls
\begin{equation}
	P_{\mathrm{dw}} \equiv \sum_{\langle i,j\rangle} \frac{1-\sigma_{i}^{z}\sigma_{j}^{z}}{2},
\end{equation}
associated with the quantum number $p\in\{0,1,\ldots,L-1\}$. Thus, $H'$ commutes with the domain-wall number $[H',P_{\mathrm{dw}}]=0$, which implies $[U_{2},P_{\mathrm{dw}}]\approx[e^{-iTH'},P_{\mathrm{dw}}]=0$.
This commutation becomes exact in the limit $V\rightarrow\infty$. Simultaneously, the spin-flip operation $U_{1}=X$ also preserves the domain wall count $[U_{1},P_{\mathrm{dw}}]=0$. Thus, the full Floquet operator approximately conserves $P_{\mathrm{dw}}$:
\begin{equation}
	[U_{F},P_{\mathrm{dw}}]=[U_{2}U_{1},P_{\mathrm{dw}}]\approx 0. 
\end{equation}
The conserved quantities $q$ and $p$ together fragment each symmetry sector $\mathcal{H}_{q}$ into finer subspaces $\mathcal{H}_{q}^{p}$ of dimension $\mathcal{D}_{q}^{p}$. This dimension can be computed combinatorially as
\begin{equation}\label{Dqp}	\mathcal{D}_{q}^{p}=a_{q}\mathcal{C}_{N_{\uparrow}-1}^{N_{\uparrow}-S_{\uparrow}}\mathcal{C}_{N_{\downarrow}-1}^{N_{\downarrow}-S_{\downarrow}}+a_{q}\mathcal{C}_{N_{\uparrow}-1}^{N_{\uparrow}-S_{\downarrow}}\mathcal{C}_{N_{\downarrow}-1}^{N_{\downarrow}-S_{\uparrow}}, 
\end{equation}
where $S_{\uparrow}=\left\lceil(p+1)/2\right\rceil$ and $S_{\downarrow}=\left\lfloor(p+1)/2\right\rfloor$ represent the number of contiguous spin-up and spin-down segments, respectively. In a word, for strong ZZ interaction $V \gg J$, the absolute magnetization and the domain-wall count serve as two approximately conserved quantities, which lead to the emergence of Floquet HSF and the constrained dynamics. The Floquet HSF provides a disorder-free mechanism of protecting the DTC order.

\begin{figure}[tb]
	\centering
	\includegraphics[width=0.48\textwidth]{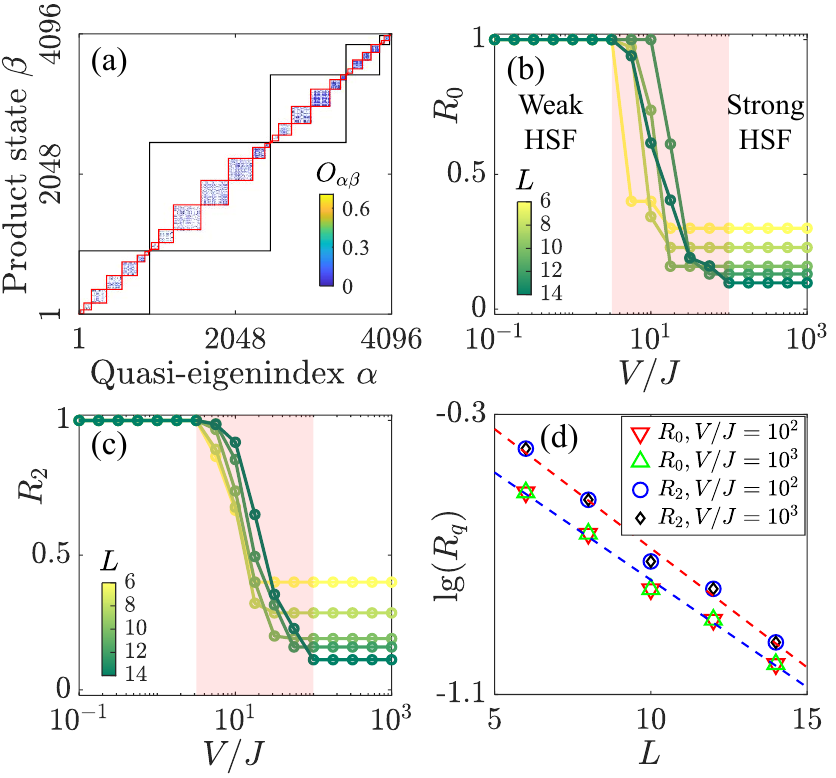}
	\caption{(Color online) Characterization of Floquet HSF with $\epsilon=0$. 
	(a) Overlap matrix $O_{\alpha\beta}$ for $L = 12$, $V/J = 10^3$, with symmetry sectors $\mathcal{H}_q$ (black) and fragmented subspaces $\mathcal{H}_q^p$ (red) outlined. 
	(b,c) Dimension ratio $R_0$ and $R_2$ as a function of $V/J$ for $L = 6\sim14$. The shaded region denotes the weak-to-strong HSF crossover. 
	(d) Exponential decay of $R_0$ and $R_2$ with $L$ for $V/J = 10^2, 10^3$. 
}
	\label{fig4}
\end{figure}

The fragmentation of the Hilbert space of the Floquet Hamiltonian $H_F$ into subspaces $\mathcal{H}_q^p$ for $L = 12$ and $V/J = 1000$ is visualized in Fig.~\ref{fig4}(a). The figure displays the overlap $O_{\alpha\beta} = |\langle\beta|\psi_\alpha^F\rangle|$ between the Hilbert space basis $|\beta\rangle$ and the Floquet eigenstates $|\psi_\alpha^F\rangle$, thresholded at $O_{\alpha\beta} > J^2/V$ and rearranged to reveal the block-diagonal structure \cite{WHLi2021}. Black and red boxes highlight the partitions into sectors $\mathcal{H}_q$ and subspaces $\mathcal{H}_q^p$, respectively. The numerical result confirms the effective disconnection of the symmetry sectors and subspaces. To quantitatively characterize the Floquet HSF under different system parameters $V$ and $L$, we employ the dimension ratio~\cite{Sala2020a,Francica2023}
{\begin{equation}	R_{q}=\frac{{\tilde{\mathcal{D}}}_{q}^{\max}}{{\mathcal{D}}_{q}},\quad{\tilde{\mathcal{D}}}_{q}^{\max}=\max_{p}\left[{\tilde{\mathcal{D}}}_{q}^{p}\right].
\end{equation}
Here, $\tilde{\mathcal{D}}_q^p$ are numerically extracted and approximate the exact $\mathcal{D}_q^p$ for $\epsilon=0$.} This ratio measures the relative size of the largest subspace within the symmetry sector $\mathcal{H}_q$. In the
thermodynamic limit, $R_q\to0$ signifies strong HSF, whereas $R_q\to1$ for weak HSF~\cite{Sala2020a,Francica2023}. Figures \ref{fig4}(b) and \ref{fig4}(c) show $R_{0}$ and $R_{2}$ as functions of $V/J$ for several system sizes, respectively. For weak ZZ interaction$ (V/J\lesssim2), R_0\approx R_2\approx1$ for all $L$s indicates the weak HSF. 
As {the interaction increases up} to $V/J\gtrsim 10$, both ratios drop sharply and saturate at values that are much less than unity and decrease with $L$. The large-$V$ behavior is further analyzed in Fig.~\ref{fig4}(d), where {$\lg(R_0)$ and $\lg(R_2)$} exhibit exponential decay with the system size $L$. 
This confirms the strong HSF with exponential suppression of heating in the thermodynamic limit~\cite{Sala2020a,Francica2023}. 
The crossover between the weak and strong HSFs is shaded in {Figs.~\ref{fig4}(b) and \ref{fig4}(c)}. 
Finally, we note that any finite perturbation $(\epsilon\neq0)$ explicitly breaks the conservation of $Q$ and $P_\mathrm{dw}$, which results in weak couplings between subspaces and eventually leads to thermalization in the long-time limit.

\subsection{Rigidity and phase regime of the DTC}
To distinguish the DTC order phase from the thermal phase, we employ the Floquet-state-averaged quantum mutual information between the two edge spins \cite{Yao2017,Amico2008a}
\begin{equation}
	\bar{M}(\mathcal{A},\mathcal{B}) = \bar{S}_F(\mathcal{A}) + \bar{S}_F(\mathcal{B}) - \bar{S}_F(\mathcal{A}\cup\mathcal{B}),
\end{equation}	
where $\bar{S}_F(\mathcal{A}) = -\frac{1}{D}\sum_{\alpha=1}^{D}\text{Tr}[\rho_{\mathcal{A},\alpha}^{F} \log \rho_{\mathcal{A},\alpha}^F]$ is the average entanglement entropy of the reduced density matrix $\rho_{\mathcal{A},\alpha}^F=\text{Tr}_{\mathcal{B}}[\ket{\psi^F_{\alpha}}\bra{\psi^F_{\alpha}}]$ for subsystem $\mathcal{A}$, and $(\mathcal{A}, \mathcal{B}) = (1, L)$ denote the first and last spins. 
The behavior of the mutual information, which measures total inter-subsystem correlations \cite{Amico2008a}, is dramatically different in the two phases. 
It approaches zero in the thermal phase as the system reaches a featureless state, but remains finite in the DTC phase due to its persistent long-range order, thus providing a clear distinction between them~\cite{Yao2017}.

\begin{figure}[tb]
	\centering
	\includegraphics[width=0.48\textwidth]{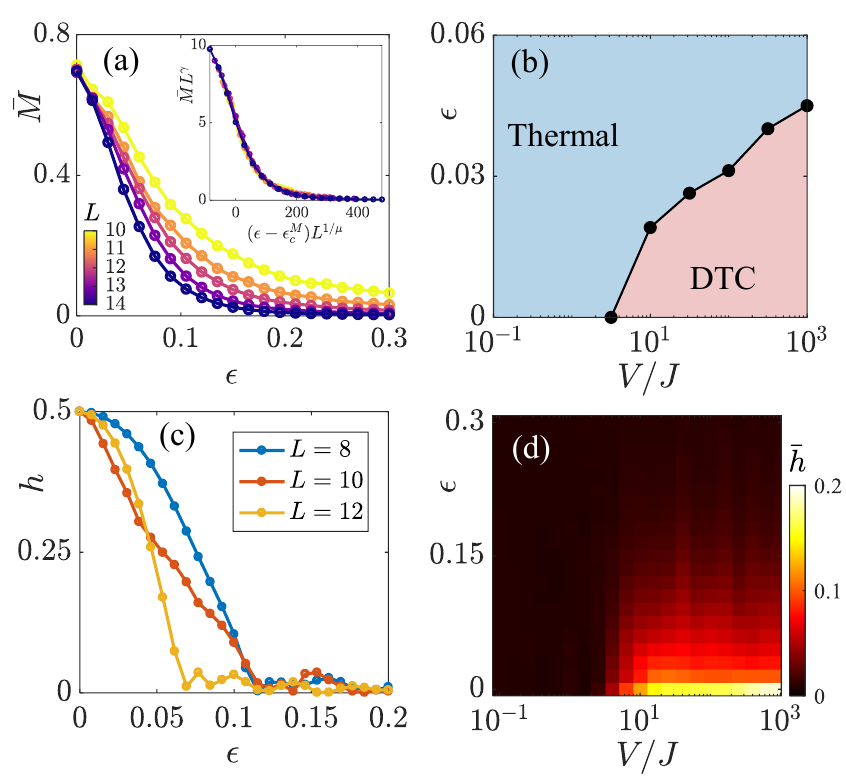}
	\caption{(Color online) 
	(a) Floquet spectrum-averaged mutual information $\bar{M}$ as a function of $\epsilon$ for $L = 10\sim14$ and $V/J = 10^3$. The inset shows the finite-size scaling collapse near the critical point $\epsilon_c^{\rm M} \approx 0.04$. 
	(b) Phase diagram in the $(V/J, \epsilon)$ plane, with dots marking the extracted phase boundaries $\epsilon_c^{\rm M}$ for different $V/J$. 
	(c) Subharmonic amplitude $h$ for the N\'eel state versus $\epsilon$ at $V/J = 100$. 
	(d) Mean subharmonic response $\bar{h}$ averaged over all initial product states   in the $\mathcal{H}_0$ sector for $L=12$.}
	\label{fig5}
\end{figure}

We study the rigidity of the DTC order with respect to the perturbation $\epsilon$ in the driving. To do this, we numerically compute $\bar{M}$ across the whole Floquet spectrum as a function of $\epsilon$ for system sizes $L = 10\sim14$ and $V/J = 10^{3}$, as shown in Fig.~\ref{fig5}(a). As $\epsilon$ increases, $\bar{M}$ decays from a finite value to zero, indicating the loss of long-range order. To precisely locate the critical point of melting the DTC, we perform a finite-size scaling analysis using the ansatz~\cite{Yao2017}
\begin{equation}
	\bar{M}\sim L^{-\gamma}f\left((\epsilon- {\epsilon_c^{\rm M}} )L^{1/\mu}\right),
\end{equation}
where the $\epsilon_c^{\rm M}$ is the critical drive imperfection, $\gamma$ and $\mu$ are two exponents. The data collapse, plotted in the inset of Fig.~\ref{fig5}(a), yields the $\epsilon_c^{\rm M}\approx0.04$, $\gamma\approx 1.5$ and $\mu\approx 0.25$. By repeating this procedure for different ZZ interaction strengths, we map out the phase diagram in the $(V/J, \epsilon)$ plane, as shown in Fig.~\ref{fig5}(b), which delineates the DTC phase from the thermal phase.

{We note that the mutual-information boundary $\epsilon_c^{\rm M}(V)$ in Fig.~\ref{fig5}(b) is a static, spectrum-averaged probe of long-range order that distinguishes the DTC phase from the thermal phase in the finite-size scaling limit. In contrast, the lifetime-scaling boundary $\epsilon_c^\tau(V)$ in Fig.~\ref{fig3}(f) is a small-size dynamical probe of exponentially protected subharmonic response that is defined and calculated within the DTC phase. These two boundaries do not coincide: the former increases with $V/J$, while the latter decreases rapidly. This complementarity highlights two distinct aspects of the DTC order enabled by Floquet strong HSF: the existence of long-range temporal order versus its exponential lifetime protection.}

Note that the phase diagram in Fig.~\ref{fig5}(b) is obtained from the mutual information $\bar{M}$, which is a static spectral property of the Floquet spectrum. 
We further validate the phase regime of the DTC from the Floquet dynamics. 
We numerically extract the subharmonic response amplitude $h$ for the Néel initial
state as a function of $\epsilon$ at $V/J=100$, as shown in Fig.~\ref{fig5}(c). The amplitude $h$ decreases from its ideal value $h\approx0.5$ down to $h\approx0$ as $\epsilon$ increases, signaling the melting of the DTC order. Furthermore, we compute the amplitude $\bar{h}$, averaged over all initial states in the $\mathcal{H}_0$ sector for $L=12$. The result of $\bar{h}$ across the $(V/J,\epsilon)$ parameter space is shown in Fig.~\ref{fig5}(d). It indicates that $\bar{h}\gtrsim0.1$ deep within the DTC phase, while $\bar{h}$ falls below 0.05 upon entering the thermal phase. The discrepancy in the phase boundaries between the static and dynamical measures can be attributed to the finite-size effects inherent in the finite-time dynamical calculations.

\section{CONCLUSION} \label{sec:conclusion and outlook}
 
In summary, we have revealed a long-lived DTC order in a disorder-free periodically kicked XXZ spin chain enabled by Floquet strong HSF. 
We have demonstrated the period-doubling and multi-period beating dynamics as well as the underlying $\pi$-pair structures for typical initial states in finite-size systems. 
The independence and enhancement of the DTC lifetime with the driving frequency and system parameters have been shown. We have analyzed the emergent fragmentation of the Floquet operator with dynamical constraints in our system, which provides a disorder-free mechanism of sustaining the non-equilibrium DTC order. The rigidity and phase regime of the DTC order have also been obtained from the static and dynamic properties of the system.

\begin{acknowledgments}
This work was supported by the National Key Research and Development Program of China (Grant No. 2024YFA1409300), National Natural Science Foundation of China (Grant No. 12174126), the Guangdong Basic and Applied Basic Research Foundation (Grant No. 2024B1515020018), 
Guangdong Provincial Quantum Science Strategic Initiative (Grant No. GDZX2404001 and No. GDZX2204003), the NSFC/RGC JRS grant (N\_HKU 774/21), GRF of Hong Kong(17303/23).
and the Science and Technology Program of Guangzhou (Grant No. 2024A04J3004).
{X.L. is supported by the Research Grants Council of Hong Kong (Grants No.~CityU~11300421, CityU~11304823, CityU~11312825 and C7012-21G) and City University of Hong Kong (Project No. 9610428).}
\end{acknowledgments}

\bibliography{reference}

\end{document}